# Do "altmetric mentions" follow Power Laws? Evidence from social media mention data in Altmetric.com

Sumit Kumar Banshal[a], Aparna Basu[b], Vivek Kumar Singh[c1], Solanki Gupta[c], & Pranab K. Muhuri[a]

[a] Department of Computer Science, South Asian University, New Delhi, India-110021
[b] Formerly CSIR-NISTADS, New Delhi-110012, India
[c] Department of Computer Science, Banaras Hindu University, Varanasi-221005, India

**Abstract**

Power laws are a characteristic distribution that are ubiquitous, in that they are found almost everywhere, in both natural as well as in man-made systems. They tend to emerge in large, connected and self-organizing systems, for example, scholarly publications. Citations to scientific papers have been found to follow a power law, i.e., the number of papers having a certain level of citation x are proportional to x raised to some negative power. The distributional character of altmetrics, (similar in concept to citations but with 'mentions' in place of citations such as read, like, mentions, etc.) has not been studied yet as altmetrics are among the newest indicators related to scholarly publications. Here we select a data sample from the altmetrics aggregator Altmetrics.com containing records from the platforms Facebook, Twitter, News, Blogs, etc., and the composite variable Alt-score for the period 2016. The individual and the composite data series of 'mentions' on the various platforms are fit to a power law distribution, and the parameters and goodness of fit determined using least squares regression. The log-log plot of the data, 'mentions' vs. number of papers, falls on an approximately linear line, suggesting the plausibility of a power law distribution. The fit is not very good in all cases due to large fluctuations in the tail. We show that fit to the power law can be improved by truncating the data series to eliminate large fluctuations in the tail. We also explore other distributions like the log-normal and Hooked Power Law that do not need data truncation and use the full data. The degree of fit of different distributions tried for altmetrics may not differ too much between the distributions. However, the importance of different models is that their generative processes are different. This underlies the importance of attributing the right model to a process, to identify what gives rise to a particular distribution. We conclude that altmetric distributions also follow power laws with a fairly good fit over a wide range of values. More rigorous methods of determination may not be necessary at present.

---

[1] Corresponding author. Email: vivek@bhu.ac.in

# Introduction

A power law is simply expressed by the fact that if the probability distribution of measuring a quantity x varies as the inverse power of the same variable raised to a value k, $p(x) \sim x^{-k}$, then x is said to follow a power law with exponent –k (Newman, 2007). Power laws were first noticed a century ago by Pareto in the context of income distribution (Pareto, 1896). As more and more distributions were found in diverse systems, their ubiquity, origins and implications have been debated with consistent regularity. They abound in nature as well as in the artificially created technology systems, most recently the Internet (Adamic, 2000; Faloutsos et al., 1999). Basically, any heavily tailed distribution could follow or model into the power law. (Unlike say the normal distribution that has a well-defined mean and variance, heavy tailed distributions are scale-free, which implies that they do not have well defined mean or variance.) This specific feature was observed in different disciplinary areas at different periods of time, over the last century by different people, and early power laws go by different names like the Bradford law (Bradford, 1934), Zipf's law (Zipf, 1949), Pareto's law (Chen & Leimkuhler, 1986; Pareto, 1896). All these types of distribution models are said to imitate the Lotka's Law (Lotka, 1926).

The nature of power laws ensures that small occurrences are very common and large occurrences are quite rare. Power law distributions involve various signature characteristics such as - if the full range is represented on the axes, the shape of the curve would be a perfect L. Also, if same distribution is plotted on log-log scale, then the curve is always linear in nature (Adamic, 2000). Also, the power law distribution is the only distribution that holds the scale–free property, i.e., the shape of distribution remains unchanged whatever be the scale at which it is observed (Newman, 2007). Although these properties can be discerned graphically, visual confirmation is rarely enough and the validity of the power law model fit to a particular data set has to be confirmed through statistical goodness of fit tests.

Power laws have also been observed in man-made systems and activities, the earliest observation being the Pareto's law of income distribution (Pareto, 1896), and recently in the Internet (Adamic, 2000; Faloutsos et al., 1999), the distribution of citations to scientific papers (Brzezinski, 2015; Redner, 1998, 2005; Thelwall & Sud, 2016) and social media (Asur et al., 2012), namely Twitter (Wang & Huberman, 2012) to name just a few.

In the domain of scholarly articles, citations are considered as a widely known parameter of impact assessment. With the evolution of social media platforms in scholarly publishing, a new and quick event based impact measure has emerged and studied widely, namely alternative metrics or 'altmetrics'(Priem & Hemminger, 2010; Priem et al., 2010). This relatively new and quick to accumulate impact indicator have been widely assessed and found to be related to citations. Altmetrics have interesting relationship with article properties (Banshal et al., 2019b; Didegah et al., 2018) and have attracted wide attention as the new kind of impact indicator (Barbaro et al., 2014; Melero, 2015). Furthermore, altmetrics have been found to correlate with citations in different degrees (Costas et al., 2015a; Eysenbach, 2011; Haustein, Peters, Sugimoto, et al., 2014; Peoples et al., 2016; Shema et al., 2014;

Thelwall, 2018; Thelwall & Nevill, 2018). However, it remains to be analysed in detail whether altmetrics follow power laws.

## Models in Power Laws

Due to the ubiquitous nature of power laws in nature, from stellar distributions, intensity of earthquakes, diversity of species a lot of studies have been done on power laws and several models have been used over the century of research into power laws. A model can tell us about the generative process under which a particular distribution is obtained, or satisfy a more general principle.

Models can be of several kinds. The model may be (*see* (Basu, 1992)),

1. *Empirical*, i.e. a function chosen in a way that it gives a good *empirical* fit to the observed data.

2. *Causal:* A model could also be *causal*, and set up the micro processes that give the desired macro result, i.e., the model may represent the underlying dynamics.

3. *Formal:* Finally, a model could be *formal* and be the result of some underlying principle e.g., entropy maximisation.

We have tried the following models, which may all be characterised as empirical models. The truncated power law, the lognormal, and the hooked power law (Thelwall, 2016a, 2016c). Since the altmetrics data is discrete, while the lognormal is for continuous data we have used the Discretized lognormal distribution and tested them for goodness of fit against data taken from Altmetric.com.

***Lognormal Distribution*** is a continuous probability distribution that fits data when the natural logarithm of the data follows a normal distribution. Since natural logarithm cannot be calculated for zero or negative numbers, hence only positive values are allowed for such distributions. The lognormal distribution is described as,

$$f(x) = \frac{1}{x\sigma\sqrt{2\pi}} e^{-(\frac{\ln(x-\mu)^2}{2\sigma^2})} \tag{1}$$

It has two standard parameters, the mean (μ) and standard deviation (σ). Modifications are required in order to convert it into discrete form for discrete data. An offset 1 is added to the mentions for such calculation giving the ***discretized lognormal distribution***.

***Hooked power law distribution.*** There are several other methods of conversion from continuous to discrete distribution (Thelwall, 2016c). **The *hooked power law*** is defined as,

$$f(x) = A * \frac{1}{(B+x)^\alpha} \tag{2}$$

where an *ad hoc* parameter B is added to the variable x. Here the two parameters that affect the process are "A" and "B". Setting the offset parameter "B" to zero will lead to the pure form of the power law.

*Truncated Power Law :* A better fit to data can be obtained by simply truncating the data series at the tail end. Again the process is ad hoc. Since the tail of the distribution is sparsely populated, the data loss may not be high. However since the fluctuations are high in the tail region, by truncation we effectively reduce fluctuations and obtain a better model fit to data.

Various multiplicative models are also used to generate such distributions. Only a small change from the generative process of one, leads to the other distribution. The power law gives a good visual fit but when logarithms are used the relationship between the raw variables may not be as accurate.

## Research Question

*Motivation:* Altmetric counts are generated from informal records on blogs, Facebook, Twitter, news, etc. by readers, as compared to the more rigorous and time-consuming process of citation by authors. It is easy to see that many individuals who read a paper can write about it on social media. Only a few of these will actually author their own a paper and cite this paper. In fact, many will tweet or otherwise mention a paper without even reading it. It is to be hoped of course that authors do indeed read the papers they are citing. So, while citation is a slow process that unfolds over time, social media is much more about the 'here and now' and interest in a topic tends to peak and die out quickly. Both processes are generated by random processes and self-organizing principles such as preferential attachment (Baraba´si & Albert, 1999; Barabási, 2013; Bianconi & Barabási, 2001).

Our objective was to test if altmetric indicators also follow power laws – as seen in the case of citations, the Internet, and numerous natural and man-made systems. It is generally assumed that the power law is seen in a play of chance and probability in large systems. If the power law for altmetrics is found we may conclude that processes in social media also follow the same laws of probability in the limit of large N (N~total mentions).

The key research questions are -

1. Do altmetrics indicators (mentions) follow Power Laws?
2. Will other distributions like hooked power law and discretised lognormal distribution be plausible for this data?

## Related Work

Power law properties are ubiquitous in various natural and behavioural phenomena. It involves properties like its scale free nature which appears to be universal. Because of this it has applications in various domains like physics (e.g. sandpile avalanches), economics, linguistics, biology (e.g. species extinction), finance, information and computer science, geology, social science, astronomy etc. (Clauset et al., 2009; Newman, 2007).

A lot of studies have been done regarding the emergence of power laws in diverse fields. Mitzenmacher (2004) tried to explain causes and the role of power laws in science. Also studied is the robust and universal behavior of power law distributions. Gabaix (2009, 2016) showed that power laws help in explaining many economic phenomena, including aggregate economic fluctuations. Bouchaud (2001) proved the power-law correlations in financial time series, whereas Faloutsos (Faloutsos et al., 1999) extended the use of these laws in network topologies to estimate the average neighbourhood size, and facilitate the design and the performance analysis of various protocol within networks.

Power laws are seen to appear in large, interconnected and self-organising systems. The statistics for number of visitors to sites of the World Wide Web was proposed and distribution of visitors per site observed, and it approximated a power law (Adamic, 2000; Adamic & Huberman, 2001). Further the web link structure of World Wide Web (WWW) was also described and studied 260,000 sites, each one of them representing a separate domain. They counted how many links a particular site received from other sites, and found that the resultant distribution obtained from links followed a power law.

There is a long list available as a proof of occurrence of power laws in various disciplines. Some of them mentioned in Newman (2007) are- word frequency count in texts, citations of scientific papers, web hits, copies of books sold, telephone calls, magnitude of earthquakes, diameter of moon craters, intensity of solar flares, intensity of wars, wealth of the richest people, frequencies of family names, population of cities and so on.

The fit of data to a power law distribution is determined by either least squares regression, or a statistical approach that combines maximum likelihood fitting with a goodness-of-fit test based on Kolmogorov-Smirnov test (Clauset et al., 2009).

The first mention about the existence of power laws in the scholarly articles was made by Solla Price (Price, 1965) for the highly cited articles. Later on, he recommended 'cummulative advantage' mechanism which also can cause power law distribution (Price, 1976). Redner (1998) analyzed the citation distribution of scientific publications and found that the asymptotic tail of the citation distribution appears to be described by a power law, with exponent equals to 3. In another work, Redner analysed articles published over a century in the journal of physical review and found the tailed characteristics of citations (Redner, 2005). Though, the exponent factor of highly cited papers and relatively lesser cited papers found to vary as studied by (Peterson et al., 2010). They analyzed two types of mechanisms in citations namely 'direct' and 'indirect' mechanism. In this approach, three different set of scientific articles were examined. The existence of power laws in citations are relatively universally found in the context of research discipline (Radicchi et al., 2008).

In a more comprehensive and elaborative empircal approach, Brzezinski (2015) detected power-law behaviour in the citation distribution. In this work, scientific papers that were published between 1998 and 2002 drawn from Scopus were analysed. They found that the power-law hypothesis is not satisfied for around half of the Scopus fields of science.

A more robust and quantitative analysis of statistical distributions was performed by Thelwall (2016c) by considering data from 26 Scopus subject areas and seven years including 911,971 journal articles. Thelwall considered three different models- the Hooked Power law model, the Truncated Power Law model, and the Discretised log normal model. He tested whether the discretised lognormal and hooked power law distributions were plausible for citation data. It also tested if there were too many uncited articles, and zero inflated variants of the discretised lognormal and hooked power law distributions (Thelwall, 2016b). He examined best options for modelling and regression, and tried to detect which distribution best fit citation data by including and excluding uncited articles (Thelwall, 2016a). He concluded that the hooked power law and discretised lognormal distribution were the best options for complete citation data. Good fits are also obtained with truncated data. Apart from citations, various other properties of scholarly article were found to follow the similar kind of distribution, namely Lotka's Law (e.g., author productivity, distribution across journals, frequencies of words occurred, average number of authors and many others (Egghe, 2005)).

The 'Altmetrics', derived from alternative metrics, are similar to citations, but are instead computed from the number of mentions of any book or paper, a scholarly document, obtains over a period of time. Similarly, it is possible to record contributions like mentions in blogs, in news or tweets. All of this can happen in a very short time as compared to citations, which usually require a person to read a paper, incorporate it in his research, and publish, all of scholarly literature is now obtained online or on the web, it is possible to record a look or a like, a read, a recommendation which taken together can take more than a year (Fang & Costas, 2020; Priem & Hemminger, 2010). Apart from the quick impact generation, altmetrics are found to follow similar kind of relationships with article properties as like as citation in various contexts. As for example, in disciplinary variation across the altmetric mentions (Banshal et al., 2019b; Sugimoto et al., 2017; Zahedi et al., 2014), different article level factors like authorship, impact factor, accessibility etc. (Didegah et al., 2018; Holmberg et al., 2020). More interestingly, altmetric data of various fields and various times have been found to be correlated with citation in different degrees. This research direction has been explored exhaustively for different disciplines Ecology (Peoples et al., 2016), Library and Information Sciences (Sotudeh et al., 2015; Thelwall & Kousha, 2017), Ornithology (Finch et al., 2017), Health and Professional Education (Maggio et al., 2018), Evolutionary Biology (Zhang & Wang, 2018) etc. In the same context two most widely used platforms namely Twitter and Mendeley have been found to be have more higher degree of correlation with citation than others (Eysenbach, 2011; Mohammadi et al., 2016; Thelwall, 2018; Thelwall & Nevill, 2018). More precisely, altmetrics found to do have a positive correlation with citations in different disciplines, journals and platforms (Costas et al., 2015a; Haustein et al., 2014).

In spite of the parallels between citations and altmetrics, the distribution of the latter has as yet not been checked. This fact induced us to try and see if the individual social media measures from Facebook, Twitter, Blogs, News and the composite variable Alt-score have data that satisfy a power law distribution.

## Data Collection

The scholarly article data has been obtained from the Web of Science (WoS)[2] for this work. The respective social media mentions have been obtained from Altmetric.com[3] aggregator. Altmetric.com accumulates social and online mentions of research output over 18 different social and online platforms such as Twitter, Facebook, Blog Mentions, News etc.

The research output of the world indexed in WoS for the year of 2016, comprising of 2,528,868 records, was downloaded. The download was performed in the month of Sep. 2019. Out of 2,528,868 records, only 1,785,149 publication records were found with DOI (Digital Object Identifier). Our focus was to collect altmetric mentions of publication data, and the DOI field provided a link between the altmetric and the bibliometric data for publication records. Therefore, only records with DOI were considered for further processing. A DOI based lookup was conducted in the Altmetric.com for the altmetric data. Out of 1,785,149 publication records, a total of 902,990 records were found to be covered by Altmetric.com, which is about 50.58% of the total data (Table 1). For all these records, the altmetric data was collected. The data obtained from Altmetric.com had 46 fields, including DOI, Title, Twitter mentions, Facebook mentions, News mentions, Altmetric Attention Score, OA Status, Subjects (FoR), Publication Date, URI, etc. The data from Altmetric.com was downloaded in the month of Sep. 2019. Out of these 902,990 papers, 891,596 records were found with at least one kind of mention.

**Table 1: Data collected from Web of Science and Altmetrics.com**

| Total WoS records (2016) | WoS records with DOI | Altmetric Records | Altmetric Records With at least 1 mention |
|---|---|---|---|
| 2,528, 868 | 1,785,149 | 902,990 | 891,596 |

## Results

In order to understand the distribution of the data, the four prominent platforms in Altmetric.com data along with the composite score have been explored, namely, Facebook, Twitter, Blogs, News and Alt-score (a combined index created out of all the other indicators). For answering the first research question posed, the standard strategy for revealing a power laws is to draw a size-frequency plot, i.e., to plot the number of papers N(k) with a given level of mentions k. In other words, we use the fact that a histogram of a quantity with a power law distribution appears as a straight line when plotted on a logarithmic scale (Newman, 2005).

The mentions were plotted using log-log plot and the parameters of fit calculated to ascertain the plausibility of a power-law. We find that the altmetrics data when plotted show reasonably good linear fit (Fig. 1). The power ranges between -1.661 to -2.614, and the variance squared explained by the model ranges between $0.8769 < R^2 < 0.9683$.

---

[2] https://www.webofknowledge.com
[3] https://www.altmetric.com/explorer

*Truncated Power Law*: It is easy to see from the plots (Fig. 1) that the fit may look good, but the parameters are not too high ($R^2 <\sim 1$). $R^2 = 0.9682$ for Blogs which gives the best fit to a power law. The least value of $R^2$ is 0.8769 for Twitter with poorest fit (Table 2).

**Table 2. Parameters of Fit to a power law and truncated power law for altmetric data**

| Data Source | Data points | A | B | R^2 | % change |
|---|---|---|---|---|---|
| Alt_Score1 | 757784 | 444464 | -1.758 | 0.9318 | - |
| Alt_Score2 | 757687 | 577279 | -1.803 | 0.9426 | 1.2 |
| Twitter 1 | 700985 | 144009 | -1.661 | 0.8769 | - |
| Twitter2 | 700714 | 1E+06 | -2.079 | 0.9614 | 9.6 |
| Facebook1 | 185243 | 69049 | -2.167 | 0.9109 | - |
| Facebook2 | 185167 | 260409 | -2.555 | 0.9683 | 6.3 |
| News | 98499 | 61276 | -1.818 | 0.8816 | - |
| Blog | 70387 | 63349 | -2.614 | 0.9682 | - |

**Note:** In each case, supplement '1' corresponds to full data and supplement '2' corresponds to truncated data

***Can the degree of fit of the Power Law be improved?*** By truncating the distribution in the tail, it is possible to obtain a better fit. It can be seen from the plots that the fit becomes poor at high values of accumulated mentions. In other words, there are only a few papers that merit a high rate of mention, and therefore there are statistical fluctuations at this end. By truncating the distribution at the high frequency end, we get an improvement in the degree of fit, without losing too many data points.

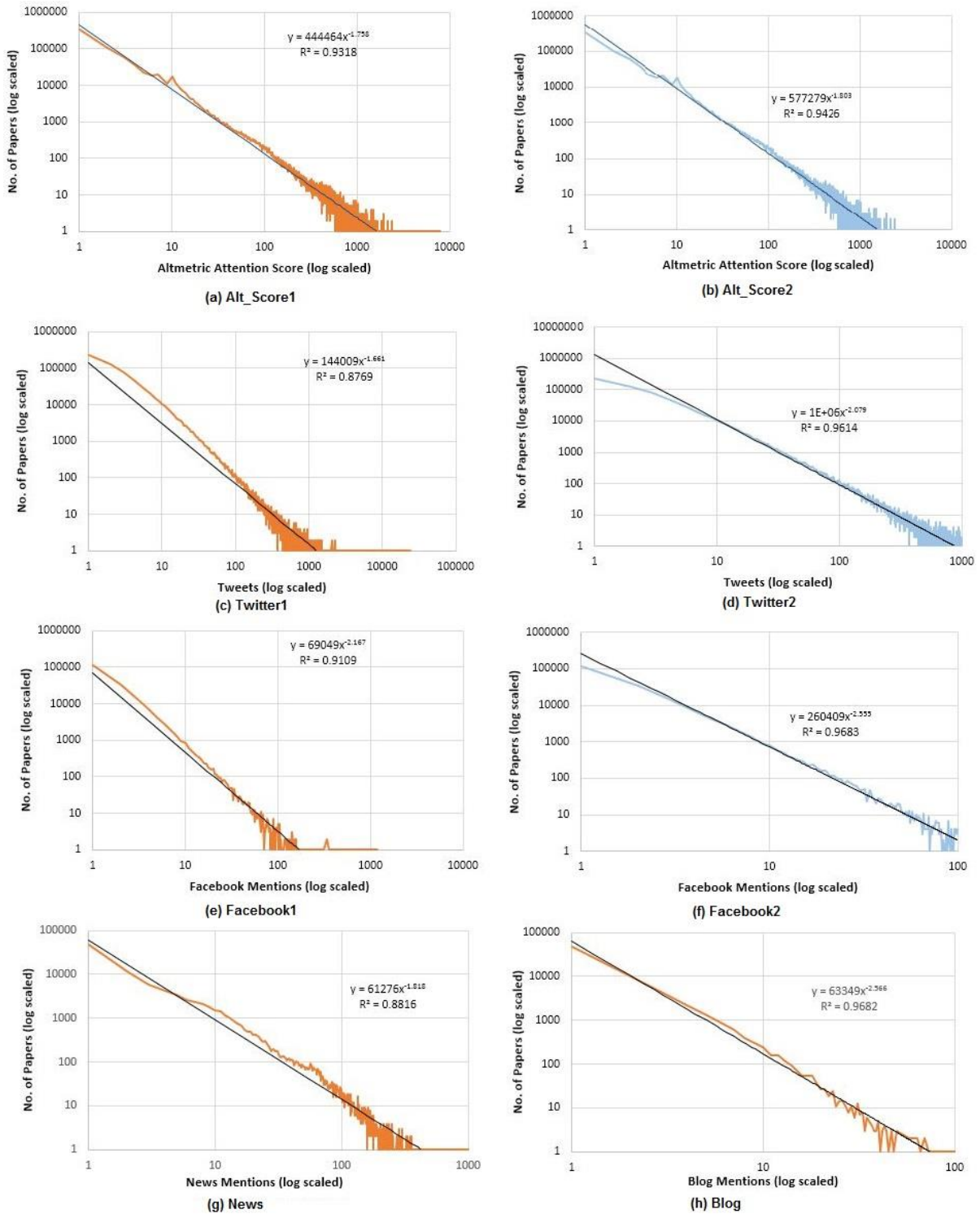

**Fig. 1 Power Law Fit to Altmetric Data: Altmetric data plotted on log-log scale with linear fit, parameters, and variance explained.**

Truncation was tried only with the Alt-score, Twitter and Facebook variables, as the data series for News and Blogs were short and therefore not suitable for truncation. The increase in the parameter of fit, $R^2$ was 1.2%, 9.6% and 6.3 %, respectively, for the Alt-score, Twitter and Facebook. (Table 2) The truncated graphs are labelled with the subscript 2 for each of the variables. (Figure 1, Table 2).

*Altmetric data fit to other distributions: Hooked Power Law and Discretized Lognormal*

As we have seen truncation of the data series can improve the goodness-of fit of the Power law model to altmetrics data. However, it can be argued that we are essentially throwing away data in order to get a good fit. Neither can it be theoretically specified which range of data to discard. There could be other distributions, other than the Power law, which give an equally good fit, or even a better fit. These distributions need to be tested against the data. We try two alternative distributions discussed by Thelwall (2016f).

An alternative to assuming that the distribution of the altmetric variables is a power law, or close to a power law, would be to look for other distributions that give a close fit to the data, this is the second research question we have noted. For addressing the second research question, the plausibility of hooked and discretised lognormal power law in altmetric data is tested. The hooked power law concept is based upon the impression that citation can occur in two processes. One of them is that citations can be accrued randomly and in the other concept, one article is getting cited due to its' earlier citations. Merging these two phenomena then the probability that an article attracts $k$ citations is proportional to $1/(B+k)^{\alpha}$. When the offset parameter, B converges to zero than it shows the distribution of pure power law (Thelwall, 2016c).

Two bootstrapped samples were drawn from all four altmetric sources (Twitter, Facebook, Blog and News) to examine the same. As mentioned by (Clauset et al., 2009) if data is compatible with power law distribution, then the data is also prone to be comparable with other empirical distributions like Discretised Lognormal, Hooked and Stretched Exponential distributions. Therefore, we checked the plausibility of hooked and discretised lognormal distributions using standard statistical techniques. The statistical approach that is followed for such calculations is stated. First, we have to consider a hypothesis based on data and then check the plausibility of the hypothesis. Given a hypothesized power law distribution from which our observed data is drawn, we are interested in finding that our hypothesis is a plausible one, given the data. We apply Kolmogorov-Smirnov (KS) statistic for estimating the *goodness of fit* for given data and calculate p-values based on KS statistic. If the p-values are larger than the level of significance, then we do not have sufficient proof to reject the hypothesis i.e. accept the null hypothesis otherwise, we reject the null hypothesis. Note that in this paper, we consider the level of significance as 0.05. To check this plausibility a code implemented earlier by Thelwall (2016c) for citation data has been used.

**Table 3. Statistical values of Hooked Power law and lognormal distribution for altmetric data**

|  | Hooked Power law | | | | Lognormal Distribution | | | |
|---|---|---|---|---|---|---|---|---|
|  | A | B | ks | p value | μ | σ | ks | p value |
| Twitter | 2.235 | 6.709 | 0.032 | 0.481 | 1.701 | 1.525 | 0.040 | 0.266 |
| Facebook | 80.566 | 208.293 | 0.316 | 0.000 | 0.990 | 0.494 | 0.151 | 0.000 |
| Blog | 99.990 | 240.950 | 0.335 | 0.000 | 0.942 | 0.468 | 0.176 | 0.000 |
| News | 2.789 | 6.031 | 0.224 | 0.000 | 1.360 | 1.078 | 0.164 | 0.000 |

* (Based on software used in (Thelwall, 2016c)}

## Discussion

The definitive identification of power law characteristics in any natural or human-made systems can be a little tricky. The phenomena has been widely explored for diverse data from many disciplinary domains, including citations, which are the most prevalent and useful indicator of the scholarly article (Brzezinski, 2015; Peterson et al., 2010; Radicchi, Fortunato, & Castellano, 2008; Thelwall, 2016a, 2016c). It had also been discussed that the nature of altmetric data and analytical tools needed for the same are quite similar to citations (Thelwall & Nevill, 2018). The essential nature of altmetric data is also heavily skewed and long tailed. Therefore, it seemed likely that altmetric data would satisfy a power law. Yet, this property had not been put to the test so far. Our approach has been to investigate the plausibility of the existence of power law behaviour pattern in altmetric mentions. It is shown that truncation of the data series can improve the closeness of fit. We obtained an increase in the $R^2$ parameter from about 1 to 9% by truncation of different altmetric data series such as Twitter, Facebook and the aggregate score Alt-score (Table 2).. Subsequently, hooked power law and discretised lognormal behaviour have also been explored on bootstrapped samples (Table 3).

### *Power Law in Altmetrics*

The first step in scrutinizing the existence of power laws in altmetric data is the log-log plot for mentions across several social media platforms- Facebook, Twitter, etc. and the composite score built from all individual scores. Altmetric.com also collects data from other platforms not mentioned here, which we have not included in the study due to the sparseness of data. The plot is shown in **Figure 1**, from which, it is observed that an approximate straight line can be fit to the data plotted on a log-log plot. This is the first characteristic of power laws and verifies the presence of power laws in altmetric mentions. These plots also specify the parameters of fit. These values help in analysing the nature of the plots and also determine where the majority of the distribution of data lies. The low (<=2) value of the power-law scaling exponent reflects an infinite (or divergent) mean, whereas mean is finite/converges for exponent>2. If α<2, it means that most of the mentions lies in tail of the distribution. Further, the median exists for exponent (B) >1 and also, variance exists if exponent >3. The parameters of fit have been listed in **Table 2**.

In the table 2, value of $R^2$ indicates the variance explained by the model, high value of $R^2$ is necessary for acceptance of the power law distribution (Clauset et al., 2009). It is also a positive evidence for our first research question. Hence, we can draw the conclusion that altmetrics mentions do follow a power law. These parameters of fit provide evidence of existence and degree of fit, but do not prove that the power law is the sole model for the observed data.

Previously, citation data has been examined and explored to find that their distribution follows power law widely across different datasets and different aspects (Brzezinski, 2015; Peterson et al., 2010; Radicchi, Fortunato, Castellano, et al., 2008). Now, in this work, the proven hypothesis of the existence of power laws in altmetric mentions reaffirms that there exists a close affinity between citation and altmetric mention. Rather than only the various degrees of correlations with citations, altmetric mentions also follow a similar pattern in data distribution. This extensively explored statistical characteristic, the power law, is also present in altmetric mentions as it is in citations.

*Discretised lognormal and Hooked Power law in Altmetrics*

After having conclusive evidence of power law in altmetric mentions, our second investigative point was to check the plausibility of discretised lognormal and hooked power law in altmetrics. To do so, we have drawn samples of 500 random observations from each category of data and then tried to estimate the population parameters for the same. Next, we applied KS statistic for estimating the *goodness of fit* for given data and calculated p-values based on KS statistic for each individual category. The results are summarized in **Table 3**.

Since the p-values of Twitter for both distributions is greater than the level of significance, it indicates that we don't have enough evidence to reject the hypothesis. Hence for twitter mentions, discretised lognormal and hooked power laws are also a plausible form of power law. Whereas in the case of Facebook, blog and news mentions, the p-values are less than the level of significance, which means these mentions do not provide a good fit for hooked and discretised lognormal distributions. Note that in this paper, we consider the level of significance as 0.05. There could be several factors for such low p-values. One could be that, these distributions are better observed only above a certain minimum value. Next could be that there are always some deviations because of the random nature of sampling.

In this approach also, altmetric mentions show similar trends like citations to some extent as both discretised lognormal and hooked power law are found to be plausible for the citation data (Thelwall, 2016a, 2016c). Though, the online mediums like blogs, news and Facebook do not show enough proof of the existence of discretised lognormal and hooked power law, Twitter is showing some indication towards the plausibility of these modified forms. It is worthy of mention that, Twitter seemed to be most prominent, explored and useful altmetric platform among these four platforms in various characteristics like coverage (Banshal et al., 2019a; Haustein, et al., 2014; Haustein, et al., 2014; Thelwall et al., 2013), correlating citations (Andersen & Haustein, 2015; Eysenbach, 2011; Peoples et al., 2016; Thelwall & Nevill, 2018) , disciplinary distributions (Banshal et al., 2019b; Costas et al., 2015b; Holmberg & Thelwall, 2014) etc.

This discussion shows that altmetric data do follow power law distribution (Table 2). Truncation of some data in the tail of the distribution improves the degree of fit. Furthermore, discretised lognormal and hooked power law provide a good fit for Twitter (tweets). Other social media platforms like Facebook, blog and news mentions do not have a good fit for the same distributions. For these the power law provides the best fit. One could further explore other forms of power laws or exponential distributions, but it appears unnecessary as the degree of fit to power law is reasonably good as seen from Table 2 and data truncation.

## Conclusion

We report here on the occurrence of power laws in altmetrics, where our data on altmetrics constitute counts of mentions, reads, downloads, blogs and news items about scientific articles. While not entirely surprising, it is nevertheless interesting to see more and newer systems getting subsumed under the same law. Not only do the above categories follow power laws, but we find that the composite altmetrics index, created by combining individual altmetric indices, also follows a power law. Finally, we test some models, used by authors earlier in similar contexts, with the observed data we have gathered from Altmetric.com.

The present study draws useful information on the nature of social media data of scholarly articles, chiefly whether they follow power law distributions which have been found to be ubiquitous in many natural and man-made systems. We find that the distribution of altmetric data does follow a power law, but with variable goodness of fit for the various social media. For a closer fit, we tried using the discretised lognormal and hooked power laws. We find that to a fair degree of approximation ($R^2 > 0.9$) data from the four social media platforms and the composite data follow power laws. The degree of fit can be further improved by some data truncation in the tail.

## Acknowledgment

The authors would like to acknowledge the access provided to data of Altmetric.com by Stacy Konkiel, Director of Research Relations at Digital Science.